\documentclass[5p, sort&compress]{elsarticle} %comment this out for single column
\usepackage{graphicx}
\usepackage{amssymb}
\usepackage{epstopdf}
\usepackage[amssymb,thinqspace,thinspace]{SIunits}
\usepackage{amsmath}
\usepackage{booktabs}
\usepackage{rotating}
\usepackage{multirow}
\usepackage{psfrag}
\usepackage{adjustbox}
\usepackage{csquotes}
\usepackage{ulem}
\usepackage{float}
\usepackage{dblfloatfix} 
\usepackage{color}
\journal{arXiv}

\begin{document}
\begin{frontmatter}
\title{A scale up study on chemical segregation and the effects on tensile properties in two medium Mn steel castings}
\author[1,*]{T. W. J. Kwok}
\author[2]{C. Slater}
\author[1,3]{X. Xu}
\author[2]{C. Davis}
\author[1]{D. Dye}

\address[1]{Department of Materials, Imperial College London, Prince Consort Road, London SW7 2BP, United Kingdom}
\address[2]{Warwick Manufacturing Group, University of Warwick, Coventry, CV4 7AL, United Kingdom}
%\address[Cam]{Department of Materials Science and Metallurgy, University of Cambridge, Pembroke Street, Cambridge, CB2 3QZ, UK}
\address[3]{School of Materials, Sun Yat-Sen University, Shenzhen, 519082, China}
\address[*]{Corresponding author, email: thomas.kwok12@imperial.ac.uk}

\begin{abstract}

Two ingots weighing 400 g and 5 kg with nominal compositions of Fe-8Mn-4Al-2Si-0.5C-0.07V-0.05Sn were produced to investigate the effect of processing variables on microstructure development. The larger casting has a cooling rate more representative of commercial production and provides an understanding of the potential challenges arising from casting-related segregation during efforts to scale up medium Mn steels, whilst the smaller casting has a high cooling rate and different segregation pattern. Sections from both ingots were homogenised at 1250 \degree C for various times to study the degree of chemical homogeneity and $\delta$-ferrite dissolution. Within 2 h, the Mn segregation range (max $-$ min) decreased from 8.0 to 1.7 wt\% in the 400 g ingot and from 6.2 to 1.5 wt\% in the 5 kg ingot. Some $\delta$-ferrite also remained untransformed after 2 h in both ingots but with the 5 kg ingot showing nearly three times more than the 400 g ingot. Micress modelling was carried out and good agreement was seen between predicted and measured segregation levels and distribution. After thermomechanical processing, it was found that the coarse untransformed $\delta$-ferrite in the 5 kg ingot turned into coarse $\delta$-ferrite stringers in the finished product, resulting in a slight decrease in yield strength. Nevertheless, rolled strips from both ingots showed $>$900 MPa yield strength, $>$1100 MPa tensile strength and $>$40\% elongation with $<$10\% difference in strength and no change in ductility when compared to a fully homogenised sample.

\end{abstract}

%\begin{keyword}
%T \sep Synchrotron Radiation \sep Austenitic Steel \sep Yield Phenomena \sep Micromechanical Modeling
%\end{keyword}
\end{frontmatter}

%\linenumbers

\section{Introduction}
%\label{Intro}
Medium Mn steels are an emerging class of \textcolor{black}{steel} which has shown great potential in energy absorbing applications. A medium Mn steel of composition Fe-10Mn-1.5Al-0.2Si-0.15C was developed as part of a 3\textsuperscript{rd} Generation Advanced High Strength Steel (3GAHSS) development project by the U.S. Department of Energy. Termed as the \enquote{High Strength $-$ Exceptional Ductility} steel, it had a tensile strength of 1200 MPa, elongation of 37\% and could be used in the front or rear pillars of an automotive Body-In-White (BIW) to protect from front or rear impact \cite{Savic2018}.

The Mn content in these steels (4$-$12 wt\%) is significantly lower than high Mn Twinning Induced Plasticity (TWIP) steels (16$-$30 wt\%) and are therefore more attractive from an industrial perspective in terms of cost and ease of production. The high Mn content in TWIP steels posed many challenges to steelmakers during industrialisation efforts over the past two decades \textcolor{black}{\cite{Redeker2007,Bausch2013}}. Feasibility studies showed that Mn segregation in cast ingots led to edge cracking during hot rolling \cite{Bausch2013,Bleck2007}. It had been theorised that these problems may be avoided in medium Mn steels due to the lower Mn content. However, medium Mn steels are still relatively heavily alloyed compared to more lean steel grades such as Dual Phase (DP) steels and similar problems faced by TWIP steels may persist.

The first major problem is chemical microsegregation during casting \cite{Wietbrock2010a,Rana2019}. While the extent of segregation may not be as severe as in TWIP steels, the Mn content is still sufficiently high to be a concern. In a study on a Quenching and Partitioning (QP) steel of composition Fe-4.5Mn-1.5Si-0.3C, Hidalgo \textit{et. al.} \cite{Hidalgo2019} showed that a segregation range of 2 wt\% Mn resulted in different martensite fractions across the steel. This resulted in inhomogeneous strain gradients during tensile testing and premature failure. Liang \textit{et. al.} \cite{Liang2018} showed in another QP steel of composition Fe-3Mn-1.5Si-.25C that Mn segregation led to banding of alternate equiaxed (Mn rich) and lath (Mn depleted) type microstructures. The difference in austenite stability between grains with the two microstructures also led to strain inhomogeniety and poor ductility. 

Tight composition control in medium Mn steels is important as slight variations in austenite composition may lead to different active deformation mechanisms, \textit{i.e.} Transformation Induced Plasticity (TRIP), TWIP or TWIP$+$TRIP \cite{Kwok2019}.  The TWIP+TRIP mechanism is usually sought after in medium Mn steels as it provides the optimal balance between strength and ductility. However, the composition window where the TWIP$+$TRIP mechanism is active is usually very narrow \cite{Lee2014}. Nevertheless, Wang \textit{et. al.} \cite{Wang2019b} showed that microsegregation in a fairly lean medium Mn steel could be avoided by twin roll casting. Their steel of composition Fe-4Mn-1.8Al-0.6Si-0.3C with a cast thickness of 2.5 mm did not show significant microsegregation and was also able to demonstrate the TWIP$+$TRIP effect after final processing. 

The second problem is the retention of $\delta$-ferrite to room temperature. \textcolor{black}{$\delta$-ferrite} is the first phase to form during the solidification of medium Mn steels but is expected to transform to austenite at typical hot rolling and slab reheat temperatures ($<1280$ \degree C) \cite{Panigrahi2001,Vazquez2020}. However, $\delta$-ferrite can be stabilised to room temperature when there are excess Al or Si additions \cite{Choi2012,Sun2018}. The effect of delta ferrite varies depending on the alloy composition and desired mechanical properties. Some researchers do not consider $\delta$-ferrite to be a problem as it does not appear to have adverse effects on tensile properties in certain alloys and may even be beneficial to ductility \cite{Choi2017,Hu2018}. However, $\delta$-ferrite is usually not beneficial from a strength perspective as \textcolor{black}{it is difficult to refine the grain size of $\delta$-ferrite and} is \textcolor{black}{therefore} typically weaker than the much finer austenite and $\alpha$-ferrite matrix \textcolor{black}{\cite{Sun2017,Sun2018}}. When deformed to large strains, the plasticity mismatch between $\delta$-ferrite grains and the matrix may cause interface cracking which might result in premature failure \cite{Sun2019}. Mn segregation at the $\delta$-ferrite interface has also been shown to reduce impact properties in medium Mn steels \cite{Kim2019a}. It is therefore important that alloy development efforts consider the size, morphology and distribution of $\delta$-ferrite in medium Mn steels and how these factors may change during a transition from lab scale to full scale production material.

Building upon previous work on a 8 wt\% Mn medium Mn steel \cite{Kwok2021}, this study aims to elucidate some of the problems which may be encountered during scale up of medium Mn steels from small scale to larger scale melts with cooling rates more representative of industrial practices. In this study, two laboratory castings of different sizes and therefore different Secondary Dendrite Arm Spacings (SDAS) were produced in order to study the effect of chemical microsegregation, homogenisation duration and $\delta$-ferrite transformation on tensile properties. It is hoped that the models developed will aid in the future development and scale up of medium Mn steels. 

\begin{figure}[t]
	\centering
	\includegraphics[width=\linewidth]{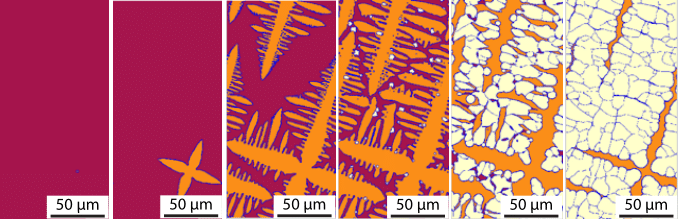}
	\caption{Solidification sequence as simulmated with Micress 6.4 and Thermo-Calc TCFE9 and MOBFE3 databases. Red $-$ liquid, orange $-$ $\delta$-ferrite, white $-$ austenite.}
	\label{fig:solidification-sequence}
\end{figure}

\section{Experimental}

Two ingots weighing 400 g (60 $\times$ 23 $\times$ 23 mm) and 5 kg (250 $\times$ 70 $\times$ 30 mm) with nominal composition Fe-8Mn-4Al-2Si-0.5C-0.07V-0.05Sn were cast using vacuum arc melting and Vacuum Induction Melting (VIM) respectively. \textcolor{black}{The choice of elements and alloying concept was described in a previous paper \cite{Kwok2021}. The 400 g ingot was prepared using pure elements and solidified in a copper mould, while the 5 kg ingot was prepared using ferroalloys and solidified in a mild steel mould. A full description of the ferroalloy compositions can be found in a previous paper \cite{Zhu2021}.} \textcolor{black}{The cooling rate of the 400 g ingot copper mould was estimated to be approximately 200 \degree C s\textsuperscript{-1} as given from the manufacturer (Arcast), while the cooling rate of the 5 kg mild steel mould was determined to be approximately 0.5 \degree C s\textsuperscript{-1} from previous work on DP steels \cite{Zhu2021}.}

\begin{figure}[t]
	\centering
	\includegraphics[width=\linewidth]{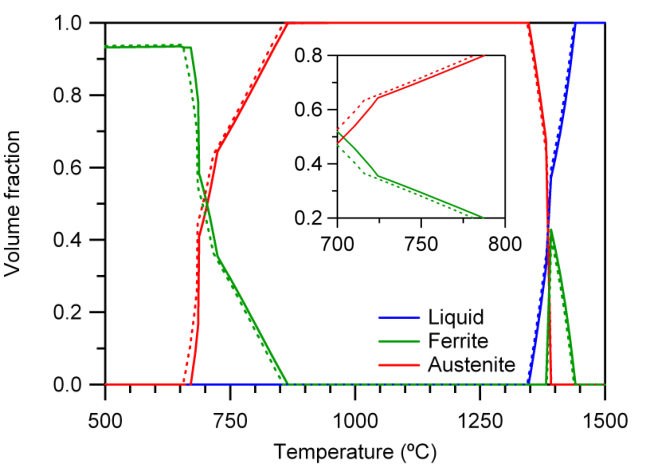}
	\caption{Thermo-Calc property diagram of the \textcolor{black}{400 g and 5 kg casting compositions as shown in Table \ref{tab:bulk composition}} using the TCFE7 database. Only liquid, ferrite and austenite phases are shown for the sake of clarity. \textcolor{black}{Solid lines $-$ 400 g ingot, dashed lines $-$ 5 kg ingot. Inset: magnified view of the property diagram at around 750 \degree C.}}
	\label{fig:nov5athermocalcaustferliq}
\end{figure}

Bars with dimensions 40 mm $\times$ 10 mm $\times$ 10 mm were cut \textit{via} Electric Discharge Machining (EDM) from each ingot, quartz encapsulated and homogenised at 1250 \degree C for 2 h before water quenching. The bars were then hot rolled to approximately 1.5 mm in 6 passes between 1000 \degree C and 850 \degree C. The hot rolled strips were immediately water quenched after rolling and Intercritically Annealed (IA) at 750 \degree C for 5 min. 

\begin{figure*}[t]
	\centering
	\includegraphics[width=\linewidth]{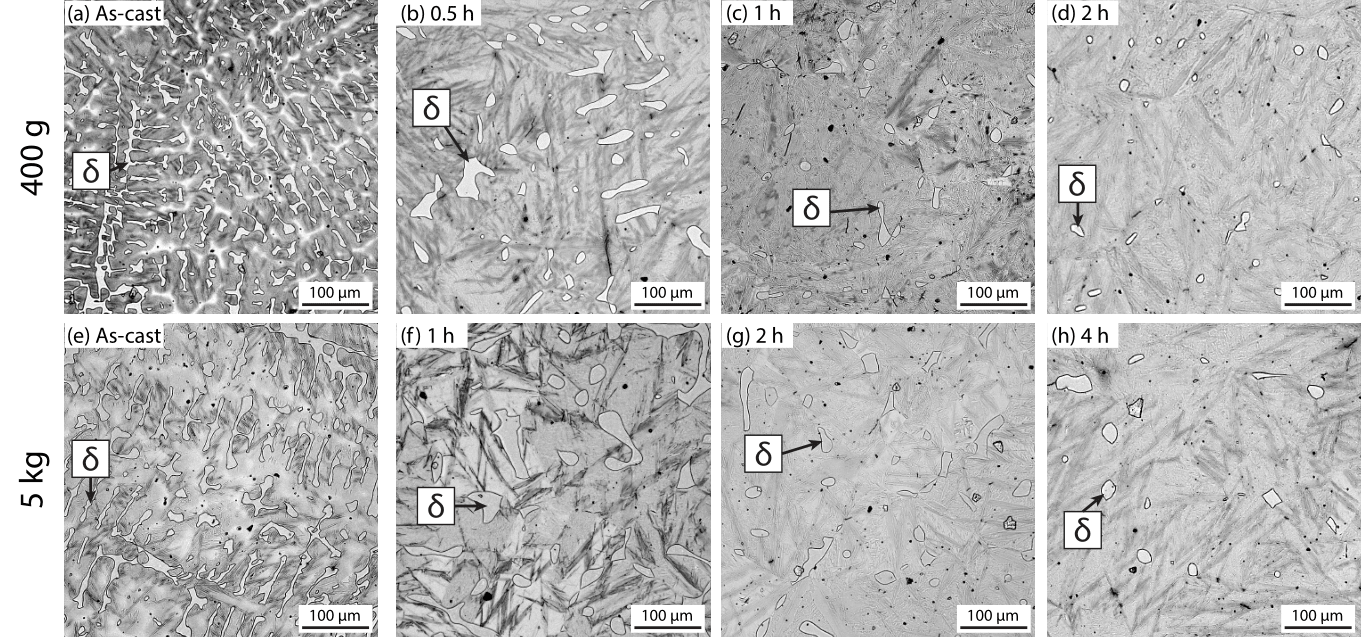}
	\caption{Optical micrographs of the 400 g ingot homogenised at 1250 \degree C for (a) 0 h, \textit{i.e.} as-cast, (b) 0.5 h, (c) 1 h, (d) 2 h and of the 5 kg ingot homogenised at 1250 \degree C for (e) 0 h, \textit{i.e.} as-cast, (f) 1 h, (g) 2 h, (h) 4 h. \textcolor{black}{The microstructures all consist of $\delta$-ferrite grains in a $\gamma/\alpha'$ matrix.}}
	\label{fig:homogenisation-opticals}
\end{figure*}

% Table generated by Excel2LaTeX from sheet 'composition'
\begin{table}[t]
	\small
	\centering
	\caption{Nominal and measured composition of the two ingots in mass percent. \textcolor{black}{Compositions measured by ICP, except for elements marked by $\dagger$ which were measured by IGF.}}
	\begin{adjustbox}{width=\columnwidth,center}
		\begin{tabular}{lccccccccc}
			\toprule
			& Mn    & Al    & Si    & C$^\dagger$     & V     & Sn    & N$^\dagger$     & P     & S$^\dagger$ \\
			\midrule
			Nom. & 8.0	& 4.0 		& 2.0		&0.5		&0.07		& 0.05		& $-$			& $-$			& $-$ \\
			400 g & 7.77      & 3.71      & 1.81      & 0.49      & 0.08      & 0.05      &  0.004     & 0.006      & 0.002  \\
			5 kg  & 8.12      & 3.46      & 1.98      & 0.46      & 0.09      & 0.05      & 0.004       & 0.026      & 0.005  \\
			\bottomrule
		\end{tabular}%
	\end{adjustbox}
	\label{tab:bulk composition}%
\end{table}%

Tensile samples with gauge dimensions of 19 $\times$ 1.5 $\times$ 1.5 mm were cut \textit{via} EDM with the tensile direction parallel to the rolling direction. Tensile testing was conducted at a nominal strain rate of $10^{-3}$ s$^{-1}$. Strain was measured with an extensometer from 0 to 10\% engineering strain and calculated from the crosshead displacement thereafter. 

Samples for microscopy were mechanically ground and polished with OP-U. Nital etchant (2\% nitric acid, 98\% ethanol) was used for etched samples. Scanning electron microscopy (SEM), Electron Backscatter Diffraction (EBSD) and Energy Dispersive Spectroscopy (EDS) were performed on a Zeiss-Sigma FE-SEM equipped with Bruker EBSD and XFlash 6160 EDS detectors. Image analysis was conducted using ImageJ software.

SEM-EDS grid scans were obtained over an area of 0.25 mm$^2$ in order to quantify the degree of segregation between the as-cast and homogenised states. 120 and 160 points in a rectangular grid were collected in the 400 g and 5 kg ingot respectively. The compositional measurements were processed according to the Weighted Interval Rank Sort (WIRS) method described by Ganesan \textit{et. al.} \cite{Ganesan2005}. The WIRS method was chosen as it was able to distinguish scatter from real element segregation trends. It first assigns a weighted value to each measurement based on the range, associated uncertainty and segregation direction of the measured element. The weighted value takes on a value between 0 and 1, where 0 represents the dendrite core and 1 represents the interdendritic region. The individual measurements are then ranked and sorted according to their weighted values.

Phase field modelling simulations were carried out using Micress 6.4 coupled with the Thermo-Calc TCFE9 and MOBFE3 databases. The model was generated using a $100\times200$ cell field with a scale of 1 $\mu$m per pixel, using an initial condition of 1530 \degree C with a single seed of delta ferrite in a liquid matrix. The seed was assigned a random location and orientation. The system was then allowed to cool at specified cooling rates of 0.5, 2, 50 and 200 \degree C s\textsuperscript{-1}, with no thermal gradient applied. These cooling rates were chosen to represent the range of solidification rates seen in different casting techniques from continuous casting, 0.6 \degree C s\textsuperscript{-1}, to lab based arc melting, 200 \degree C s\textsuperscript{-1}.

Relinearisation of the phase diagram was applied every 5 \degree C to help resolve the solute build-up at the liquid/solid interface. A nucleation condition of the FCC structure was allowed on the BCC/liquid interface at temperature below that of the equilibrium peritectic temperature (1380 \degree C). The model implements a periodic/wrap around symmetry and as such any features that extend out of the right side of the image will appear from the left (the same is true in the vertical axis). Finally, a time step of 0.1/cooling rate was applied to ensure sufficient resolution at all cooling rates. A typical solidification sequence can be seen in Figure \ref{fig:solidification-sequence}.

After conducting the solidification sequence, the output was then reapplied to Micress to apply a heat treatment. A constant temperature of 1250 \degree C and simulations of up to 4 hours were applied. Finally, the final phase balance and compositional distribution were exported.

\section{Results and discussion}

\subsection{As-cast and homogenised microstructure}
The bulk composition of the two ingots were measured using Inductively Coupled Plasma (ICP) and Inert Gas Fusion (IGF) and the results are shown in Table \ref{tab:bulk composition}. The 400 g ingot showed a slightly lower Mn content due to the high heat input from the arc melting process, leading to excessive Mn vapourisation. Apart from Mn, all other elements were comparable and within acceptable tolerances.

\begin{figure*}[t]
	\centering
	\includegraphics[width=\linewidth]{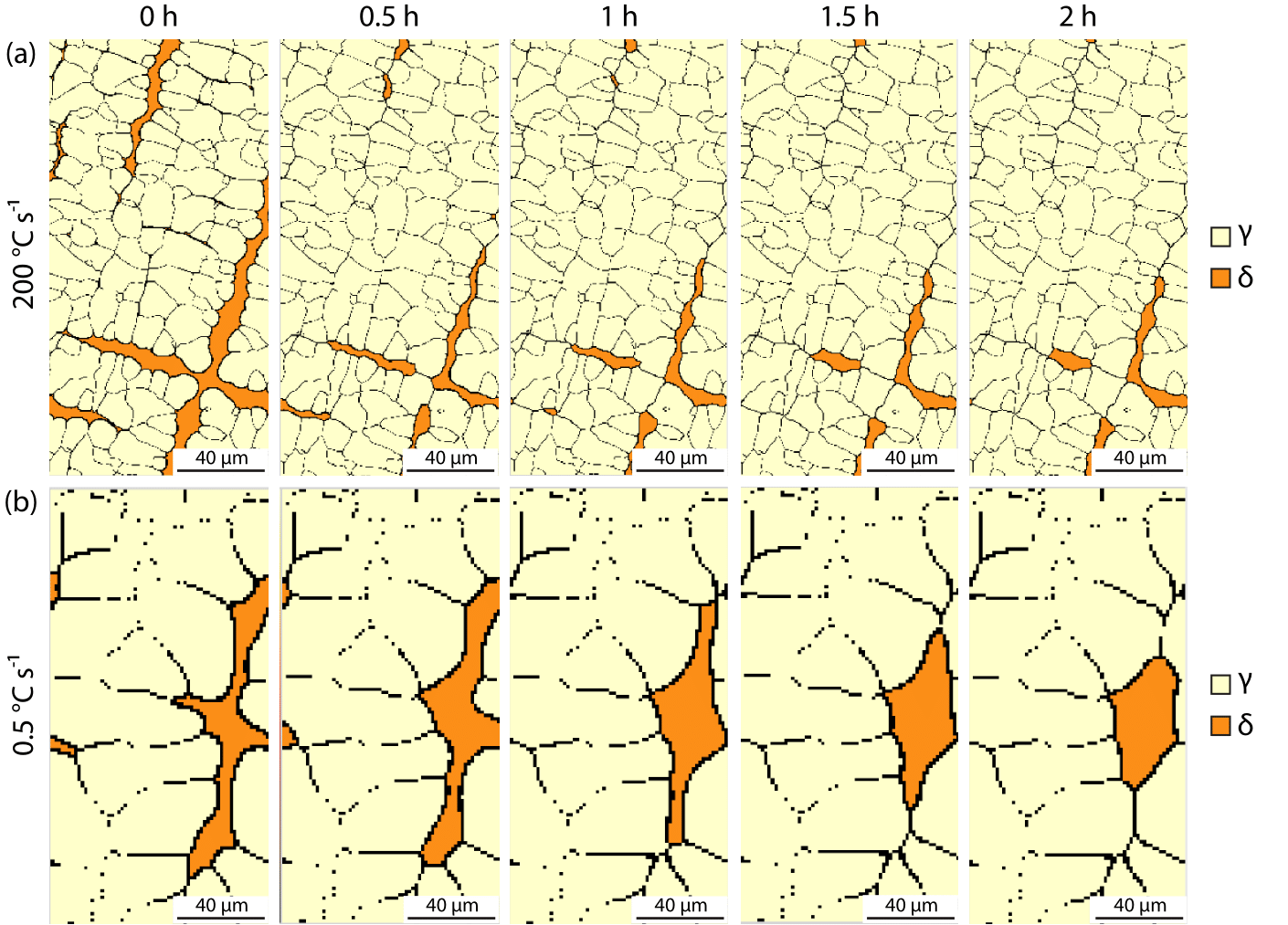}
	\caption{Micress simulation of the microstructure evolution during the homogenisation heat treatment at 1250 \degree C. Initial cast microstructures were formed from cooling rates of (a) 200 \degree C s\textsuperscript{-1}, representative of the 400 g ingot and (b) 0.5 \degree C s\textsuperscript{-1}, representative of the 5 kg ingot. }
	\label{fig:homogenisation-micress-phase-only}
\end{figure*}

Figure \ref{fig:nov5athermocalcaustferliq} shows the property diagram of the \textcolor{black}{400 g and 5 kg steel compositions} as simulated by Thermo-Calc using the TCFE7 database. \textcolor{black}{Thermo-Calc is often used during the alloy design stage to give an indication of phase fractions and approximate compositions of each phase, especially within the intercritical temperature regime. In most medium Mn steels in the literature, the phases and compositions obtained during intercritical annealing are retained to room temperature \cite{Li2020,Lee2013c,Farahani2015,Kwok2021}. From Figure \ref{fig:nov5athermocalcaustferliq}, it can be seen that the slight differences in composition between the 400 g and 5 kg ingot did not result in a significant change in the expected phase fractions at the chosen IA temperature of 750 \degree C.} \textcolor{black}{Thermo-Calc also} predicted that $\delta$-ferrite should not be expected below 1370 \degree C but it was clearly observed in the as-cast micrographs of both 400 g and 5 kg ingots in Figure \ref{fig:homogenisation-opticals}. From Figure \ref{fig:homogenisation-opticals}, the as-cast micrographs showed $\delta$-ferrite dendrite cores and an austenite/martensite matrix. The average SDAS was determined by measuring the number of secondary arms along several primary dendrite arms (method D by Vandersluis \textit{et. al.} \cite{Vandersluis2017}) and found to be $19\, \pm\, 5$ $\mu$m in the 400 g ingot and $29\, \pm\, 4$ $\mu$m in the 5 kg ingot. During homogenisation, $\delta$-ferrite was observed to spheroidise before shrinking in area. After 2 h, the $\delta$-ferrite grains in the 400 g ingot were much smaller in size, approximately 20$-$25 $\mu$m wide, while the $\delta$-ferrite grains in the 5 kg ingot were slightly larger, approximately 30$-$40 $\mu$m wide.

The observed microstructures were in good agreement with the Micress modelling for the two cooling rates as shown in Figure \ref{fig:homogenisation-micress-phase-only} where the size and distribution of the $\delta$-ferrite was clearly dependent on the solidification rate. In the model, the spheroidisation of the $\delta$-ferrite can be seen to occur during the homogenisation heat treatment with the 0.5 \degree C s\textsuperscript{-1} cooling rate showing much coarser globular $\delta$-ferrite after 2 h. 

\begin{figure}[t]
	\centering
	\includegraphics[width=\linewidth]{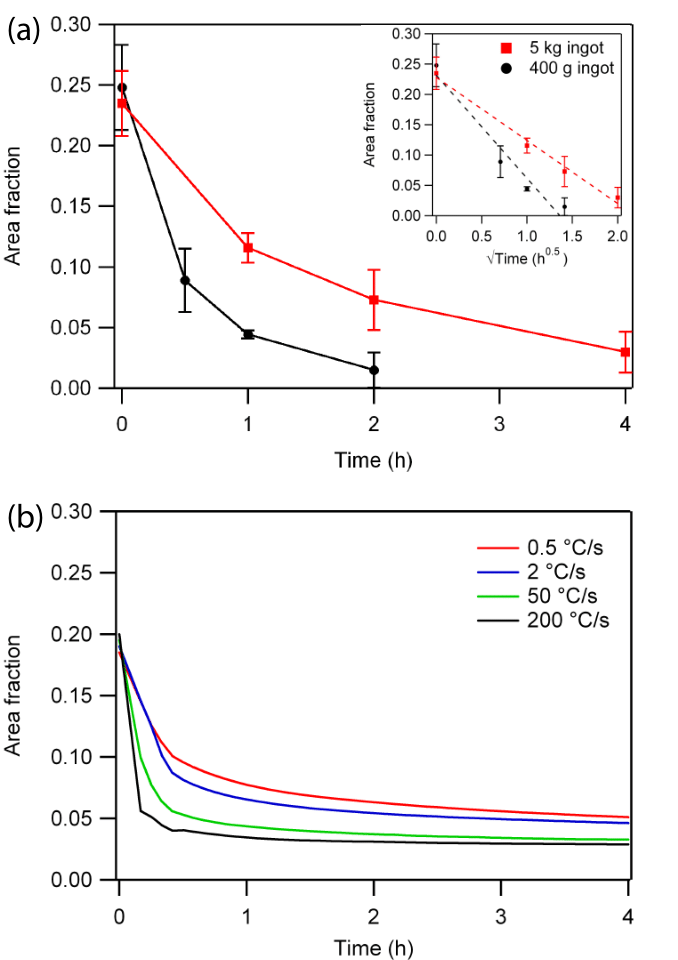}
	\caption{(a) Change in area fraction of $\delta$-ferrite with homogenisation time at 1250 \degree C and replot as a function of the square root of time in the inset. (b) Micress simulation of the change in $\delta$-ferrite fraction with homogenisation time at 1250 \degree C from ingots cast with different cooling rates..}
	\label{fig:delta-area-frac-graph}
\end{figure}

The experimentally observed $\delta$-ferrite area fraction as a function of homogenisation time is shown in Figure \ref{fig:delta-area-frac-graph}a. In the as-cast condition, both ingots had \textcolor{black}{an approximate $\delta$-ferrite area fraction of 0.25}, regardless of SDAS. Within 2 h of homogenisation, which is a typical duration in an industrial reheating furnace, the area fraction decreased to 0.02 and 0.07 in the 400 g and 5 kg ingot respectively. When $\delta$-ferrite area fraction was plotted as a function of the square root of homogenisation time, the decrease in $\delta$-ferrite exhibits a linear trend. This suggests that the transformation of $\delta$-ferrite to austenite at 1250 \degree C is a diffusion-limited process where the mean diffusion distance, $x$, in one dimension may be approximated as:

\begin{equation}
\label{eq:diffusion}
x \approx \sqrt{Dt}
\end{equation}

\noindent where $D$ is the diffusion coefficient and $t$ is the time for diffusion. 

When comparing the experimental results with the model (Figure \ref{fig:delta-area-frac-graph}b), a very good agreement can be observed. The model accurately predicts that the initial $\delta$-ferrite fraction would be similar, regardless of cooling rate. However, it predicts a slightly lower initial area fraction than what was experimentally observed. Nevertheless, the model was able to show how the rate of $\delta$-ferrite dissolution increases at higher cooling rates due to the much finer SDAS and therefore shorter diffusion distance. This highlights the need for full scale prototyping as small scale laboratory productions may under-predict the time necessary for homogenisation.

It is noteworthy that the model showed that the rate of homogenisation, \textit{i.e.} $\delta$-ferrite reduction, slows significantly at long durations ($>$ 2 h). However, it was observed experimentally that while the rate of homogenisation slowed, it was not as severe as the prediction. This was likely due to a limitation of the model being 2D and not 3D, giving one less degree of freedom for diffusion. While the impact of one less degree of freedom is negligible at short durations ($<$ 2 h), the difference becomes more significant at longer durations.

\subsection{Chemical segregation}

\begin{figure*}[t]
	\centering
	\includegraphics[width=\linewidth]{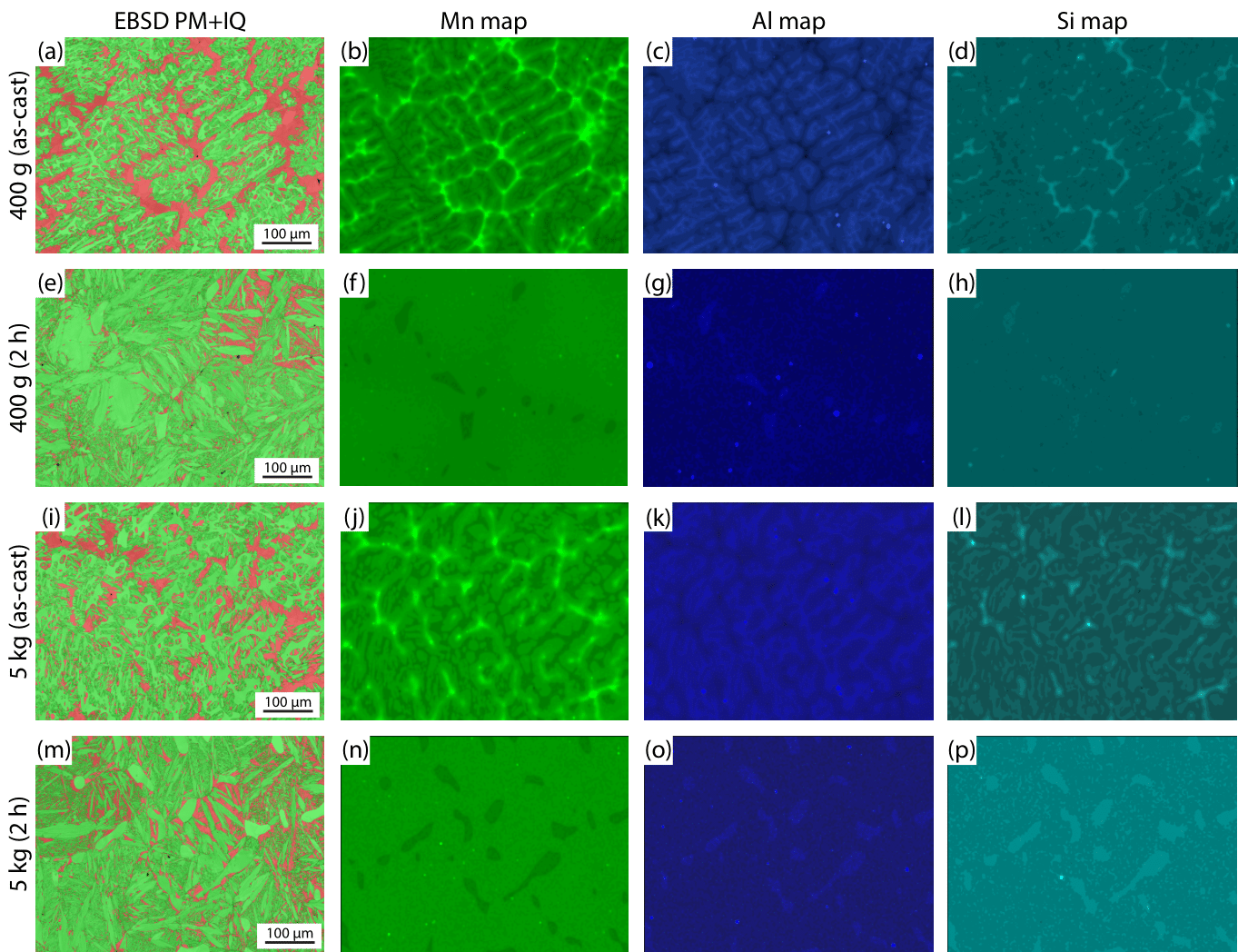}
	\caption{From left to right columns: EBSD phase maps and image quality (green $-$ BCC, red $-$ FCC), EDS-Mn, EDS-Al and EDS-Si maps of \textcolor{black}{the} as-cast 400 g ingot \textcolor{black}{(a-d)}, 2 h homogenised 400g ingot \textcolor{black}{(e-h)}, as-cast 5 kg ingot  \textcolor{black}{(i-l)} and 2 h homogenised 5 kg ingot \textcolor{black}{(m-p)}. For EDS maps, a brighter colour indicates a higher concentration and a darker colour indicates a lower concentration. }
	\label{fig:ebsd-eds-maps}
\end{figure*}

Correlative EBSD and EDS maps (Figure \ref{fig:ebsd-eds-maps}) of the as-cast and 2 h homogenised conditions of both ingots were collected for the major alloying elements: Mn, Al and Si. In the as-cast condition of both ingots, a significant fraction of austenite was present at room temperature in the form of both untransformed and retained austenite, shown in Figures \ref{fig:ebsd-eds-maps}a and \ref{fig:ebsd-eds-maps}i. Both $\delta$-ferrite and martensite which indexed as BCC phases were not as easily distinguished in the EBSD phase maps. However, the three phases can be identified with certainty from the EDS Mn maps (Figures \ref{fig:ebsd-eds-maps}b and \ref{fig:ebsd-eds-maps}j) which showed three distinct regions with low, medium and high concentrations of Mn. The low, medium and high Mn regions correlate with $\delta$-ferrite, martensite/austenite and retained austenite regions respectively. Mn was therefore observed to segregate strongly to the interdendritic regions. Al similarly showed three distinct regions but with the opposite segregation direction. The high, medium and low Al regions correlated with the $\delta$-ferrite, martensite/austenite and retained austenite regions respectively (Figures \ref{fig:ebsd-eds-maps}c and \ref{fig:ebsd-eds-maps}k). In the as-cast 400 g ingot, Si also showed three distinct regions but only the high Si regions correlated with the interdendritic regions, suggesting that Si segregates to the interdendritic region (Figure \ref{fig:ebsd-eds-maps}d). However, in the as-cast 5 kg ingot, Si appeared to segregate to both $\delta$-ferrite and the interdendritic region (Figure \ref{fig:ebsd-eds-maps}l). The difference in Si partitioning may be due to the difference in cooling rate. At fast cooling rates observed in the 400 g ingot, Si partitions to the interdendritic region, but at slower cooling rates observed in the 5 kg ingot, there would have been sufficient time for back diffusion of Si to the $\delta$-ferrite grains.

\begin{figure*}
	\centering
	\includegraphics[width=\linewidth]{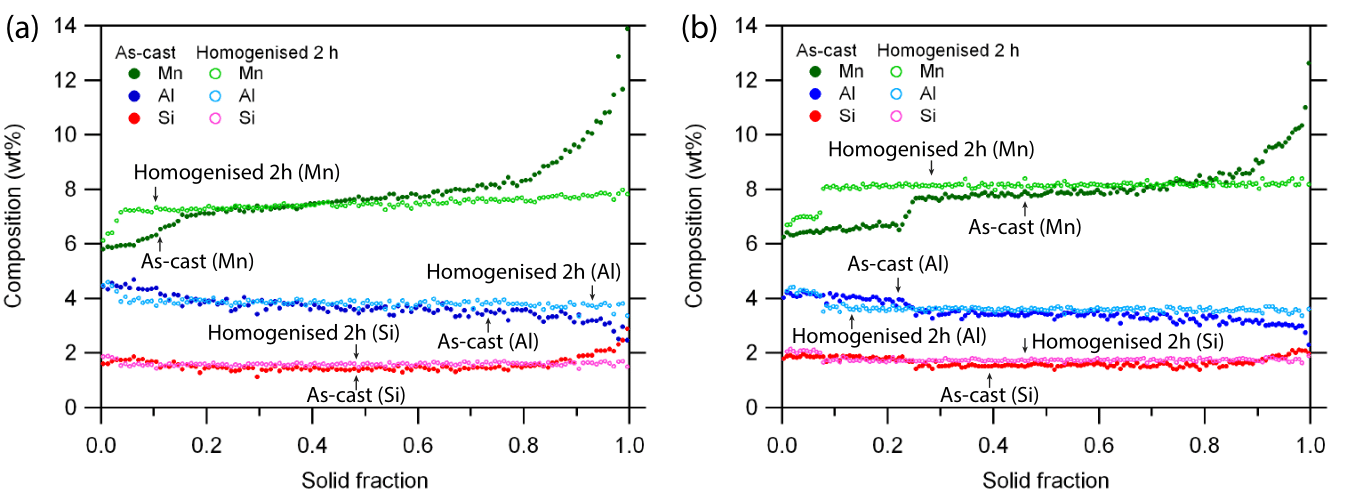}
	\caption{WIRS plots of the as-cast and 2 h homogenised conditions of (a) 400 g and (b) 5 kg ingots. A solid fraction of 0 represents the dendritic core regions and a solid fraction of 1 represents the interdendritic regions. }
	\label{fig:wirs}
\end{figure*}

After 2 h, $\delta$-ferrite was still present and easily identified as the Mn depleted regions in the EDS-Mn maps (Figures \ref{fig:ebsd-eds-maps}f and \ref{fig:ebsd-eds-maps}n). The Mn concentration of the matrix in both ingots now appear to be homogenous. The Al (Figures \ref{fig:ebsd-eds-maps}g and \ref{fig:ebsd-eds-maps}o) and Si (Figures \ref{fig:ebsd-eds-maps}h and \ref{fig:ebsd-eds-maps}p) EDS maps show that that $\delta$-ferrite was slightly enriched in Al and Si while the matrix was homogeneous. It should also be mentioned that several alumina and silica inclusions were present in both 400 g and 5 kg ingots in the as-cast and 2 h homogenised conditions.

A more quantitative WIRS plot from EDS gridscans of the same areas in Figure \ref{fig:ebsd-eds-maps} is shown in Figure \ref{fig:wirs}. In the as-cast condition of both ingots, Mn was found to have the highest level of segregation. As observed in the EDS map, Mn was found to segregate strongly to the interdendritic regions (\textit{i.e.} high solid fraction). The degree of Mn segregation was greater in the 400 g ingot and the maximum Mn content reached as high as 14 wt\%. This effect is consistent with the faster cooling rate in the 400 g ingot and therefore less time for back diffusion. At low solid fractions, a lower shelf of Mn can be observed which can be attributed to $\delta$-ferrite.

The lower Mn shelf in both as-cast ingots corresponded with a slight increase in Al concentration confirming that Al segregates to $\delta$-ferrite during solidification. Si did not segregate strongly in both ingots except in the 400 g ingot where there was a slight increase in Si at the interdendritic regions. 

After a 2 h homogenisation heat treatment, the Mn segregation at the interdendritic regions was mostly eliminated in both ingots. However, the lower Mn shelf was still present which corresponds to the remaining $\delta$-ferrite. Segregation of Al and Si at the interdendritic regions were also mostly eliminated and remained fairly constant throughout the matrix.

\begin{figure*}
	\centering
	\includegraphics[width=\linewidth]{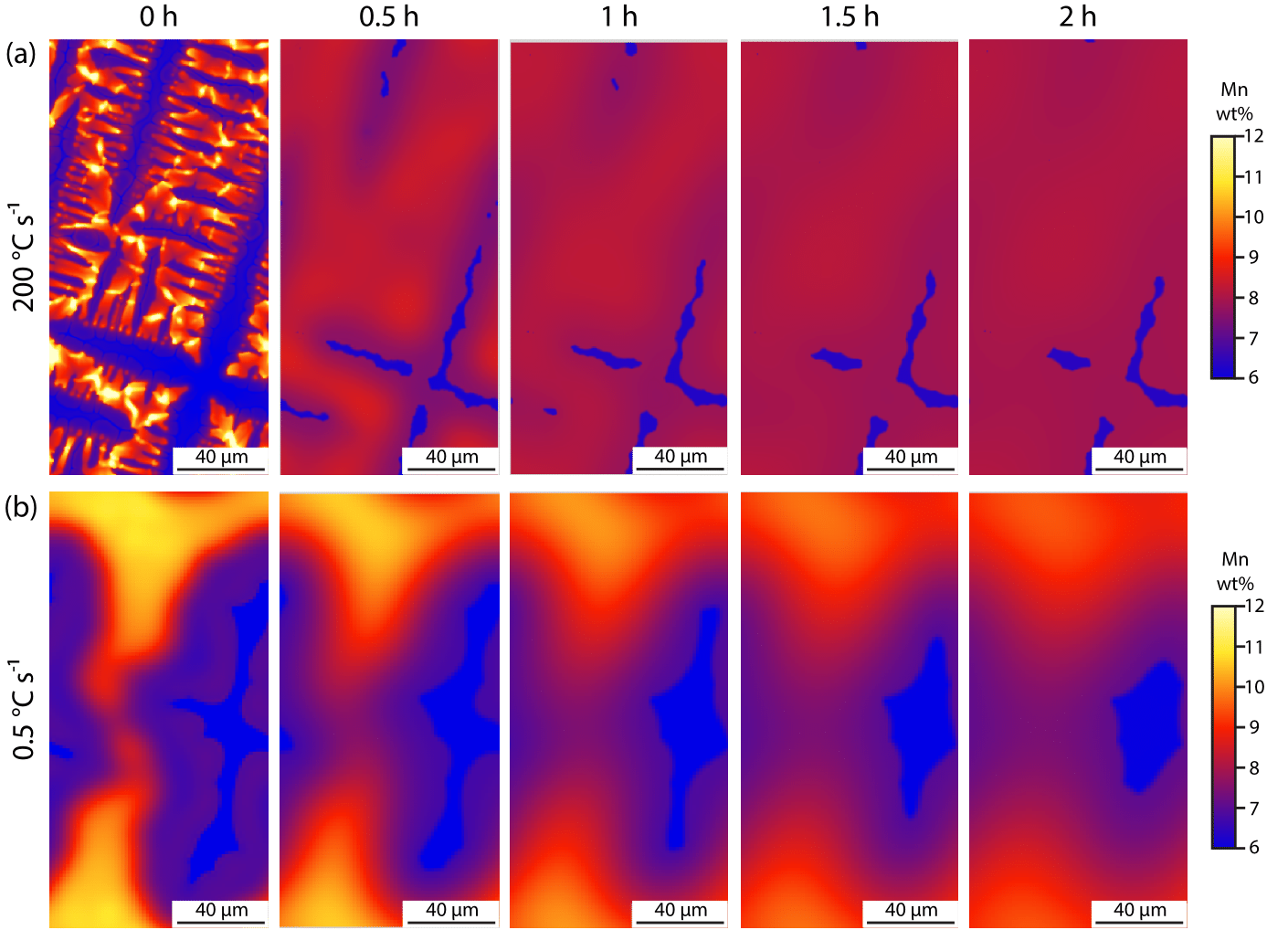}
	\caption{Micress segregation simulation during the homogenisation heat treatment at 1250 \degree C. Initial cast microstructures were formed from cooling rates of (a) 200 \degree s\textsuperscript{-1}, representative of the 400 g ingot and (b) 0.5 \degree s\textsuperscript{-1}, representative of the 5 kg ingot. }
	\label{fig:homogenisation-micress-mn-only}
\end{figure*}

Good agreement between modelling and experimental can be seen in Figure \ref{fig:homogenisation-micress-mn-only} where the $\delta$-ferrite can be seen to have a similar Mn content of approximately 6 wt\%. Very quickly during the homogenisation process, the Mn rich regions began to homogenise, particularly in the 400 g ingot, eliminating the smaller secondary and tertiary dendrite arms. 

\begin{figure}
	\centering
	\includegraphics[width=\linewidth]{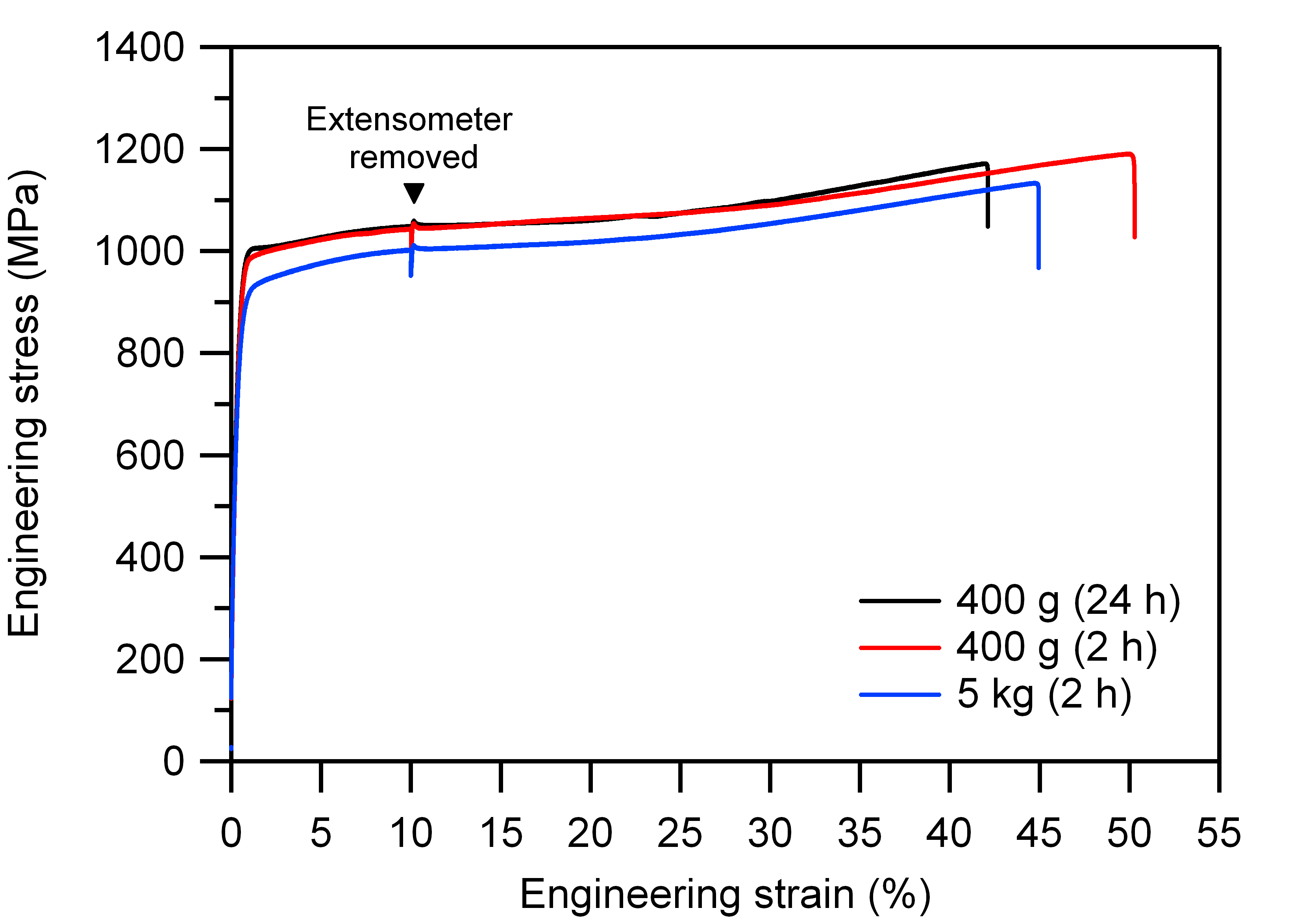}
	\caption{Tensile behaviour of the partially and fully homogenised rolled and IA samples.}
	\label{fig:nov5a2-a3-wmg-24hr-2hr-tensiles}
\end{figure}

\begin{table}[t]
	\centering
	\caption{Tensile properties of the rolled strips.}
	\begin{tabular}{lcccc}
		\toprule
		\multirow{2}[0]{*}{} & $E$ & $\sigma_{y}$ & $\sigma_{UTS}$ & $\epsilon$ \\
		& (GPa) &(MPa) & (MPa) & (\%)  \\
		\midrule
		400 g (24 h)  & 166 & 1005  & 1170  & 42  \\
		400 g (2 h)  & 175 & 985   & 1189  & 50  \\
		5 kg (2 h)  & 156 & 918  & 1133  & 45  \\
		\bottomrule
	\end{tabular}%
	\label{tab:tensile_props}%
\end{table}%

\subsection{Tensile properties and rolled microstructure}
Tensile samples were obtained from rolled and IA strips after homogenising a bar from the 400 g and 5 kg ingot for 2 h at 1250 \degree C, hereby known as 400 g (2 h) and 5 kg (2 h). The tensile curves are shown in Figure \ref{fig:nov5a2-a3-wmg-24hr-2hr-tensiles} together with a tensile curve obtained from previous work on a fully homogenised sample \cite{Kwok2021}. The fully homogenised sample, hereby known as 400 g (24 h), was produced using the same thermomechanical processing steps but from a 400 g ingot which was homogenised for 24 h. A summary of the tensile properties is shown in Table \ref{tab:tensile_props}. 

\textcolor{black}{It is acknowedged that the use of sub-sized tensile samples might introduce geometry-related errors in the measured tensile properties. Unfortunately, due to the load capacity of the laboratory rolling mill, it is not possible to roll a larger bar to obtain larger tensile samples. Nevertheless, it has been shown that the yield strength, tensile strength and uniform elongation are largely independent of sample geometry \cite{Hanlon2015}. Since none of the tensile samples showed any post uniform elongation in Figure \ref{fig:nov5a2-a3-wmg-24hr-2hr-tensiles}, it is likely that the tensile properties in the sub-sized tensile samples will be similar to a full sized sample. }

From Figure \ref{fig:nov5a2-a3-wmg-24hr-2hr-tensiles}, the tensile samples from the 400 g ingots showed very similar strengths. The 400 g (24 h) sample was slightly stronger but the 400 g (2 h) sample was more ductile. The 5 kg (2 h) tensile sample was 87 MPa lower in yield strength compared to the 400 g (24 h) sample but also had slightly more ductility. All three tensile samples showed nearly identical strain hardening behaviour, strongly indicating that the TWIP$+$TRIP deformation mechanism was active in a similar manner across the three tensile samples in spite of the slight difference in Mn as shown in Table \ref{tab:bulk composition}.

\begin{figure*}[h]
 	\centering
 	\includegraphics[width=\linewidth]{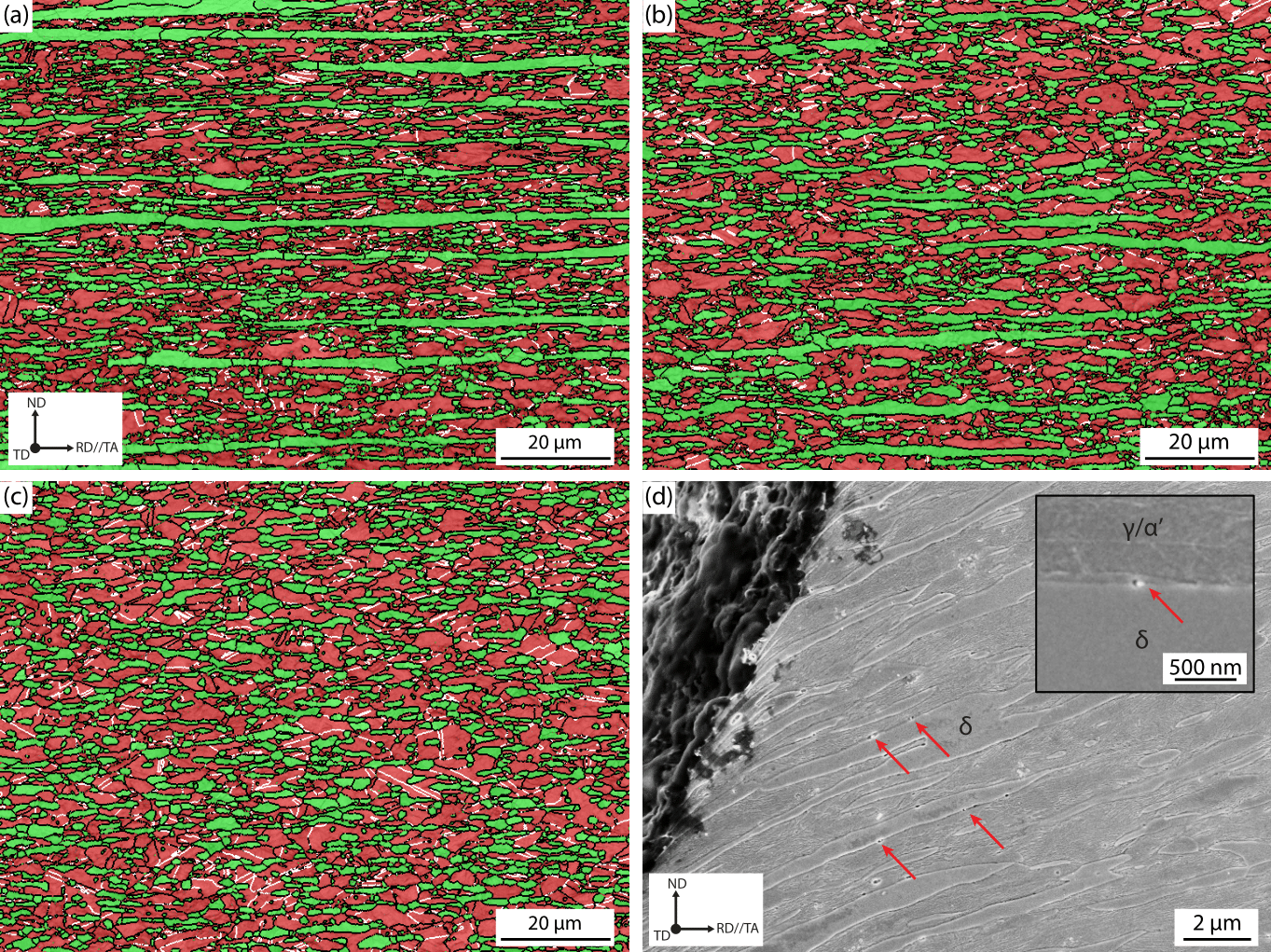}
 	\caption{EBSD image quality and phase maps of the rolled and intercritically annealed strips from the  (a) 5 kg (2 h) ingot, (b) 400 g (2 h) ingot and (c) 400 g (24 h) ingot. Red - austenite, green - ferrite. Black lines indicate grain boundaries and white lines indicate austenite $\Sigma3$ boundaries. (d) Secondary electron micrograph of a sheared edge along the fractured surface of the 5 kg (2 h) tensile specimen. Red arrows point to several nanoscale precipitates inside a void along interphase boundaries. Inset: magnified view of a precipitate along a $\delta$-ferrite stringer. }
 	\label{fig:rolled-ebsd}
 \end{figure*}

The rolled and IA microstructures of the three tensile samples are shown in Figure \ref{fig:rolled-ebsd} and the phase fractions are shown in Table \ref{tab:tensile EBSD phase fraction}. The fully homogenised 400 g (24 h) sample (Figure \ref{fig:rolled-ebsd}c) showed fine equiaxed ferrite grains, located predominantly along the austenite grain boundaries. This accurately reflects the process route where the steel was first rolled in a temperature regime where the steel was fully austenitic and the ferrite grains were newly formed $\alpha$-ferrite grains which nucleated during the final passes and the intercritical annealing heat treatment. By comparison, the 5 kg (2 h) sample showed the presence of large elongated ferrite grains which are most likely $\delta$-ferrite stringers, consistent with the large globular $\delta$-ferrite grains after homogenisation which became elongated during rolling. The 400 g (2 h) sample also showed the presence of $\delta$-ferrite stringers but were shorter compared to the 5 kg (2 h) sample due to the smaller globular $\delta$-ferrite size before hot rolling. Fine ferrite grains were also observed in the partially homogenised samples which are likely to be $\alpha$-ferrite which formed in the same way as the 400 g (24 h) sample.

The $\delta$-ferrite fraction after rolling can be estimated by subtracting the amount of newly formed $\alpha$-ferrite of the 400 g (24 h) sample from the total ferrite fraction of the 5 kg (2 h) and 400 g (2 h) rolled strips. This results in a $\delta$-ferrite fraction of 7.2\% and 4.4\% respectively, which is in close agreement with the $\delta$-ferrite fraction after the 2 h homogenisation heat treatment (Figure \ref{fig:delta-area-frac-graph}a) of 6.5\% and 3.5\% shown both experimentally and through modelling. This also shows that any remaining $\delta$-ferrite after homogenisation will be retained to room temperature even after hot rolling.

The volume fraction and distribution of $\delta$-ferrite can be seen to have a significant impact on tensile performance with an 87 MPa drop in yield strength when comparing the 400 g (24 h) and 5 kg (2 h) tensile sample. This phenomenon is common in dual phase steels where the spatial distribution of the second phase, such as band spacing, has been shown to have a noticeable impact on tensile performance \cite{Slater2020}. In this medium Mn steel, the stability of $\delta$-ferrite phase was shown to be a function of the SDAS which is determined by the casting conditions. Consequently, the size and distribution of $\delta$-ferrite in the final microstructure would be determined by both casting condition and rolling reduction. Therefore, a finer SDAS from the cast condition will give a lower $\delta$-ferrite fraction with a finer band spacing and result in a stronger steel.

\begin{figure}[t]
	\centering
	\includegraphics[width=\linewidth]{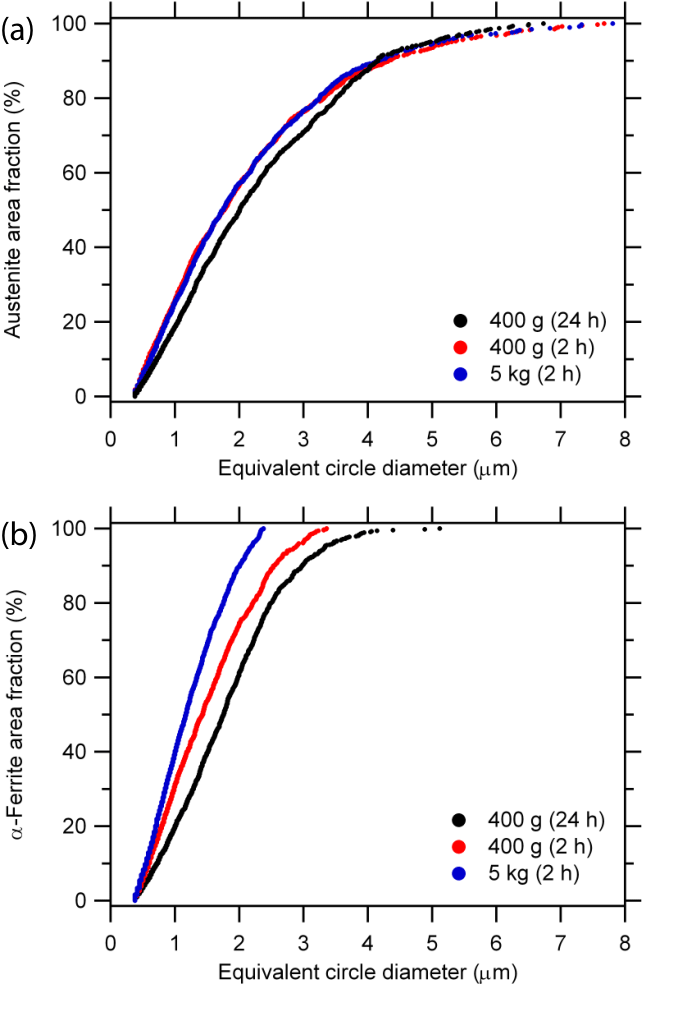}
	\caption{Cumulative distribution function of (a) austenite and (b) $\alpha$-ferrite grain sizes across the three rolled strips. }
	\label{fig:grain-size-weibull-area-frac}
\end{figure}

The relationship between casting condition and grain size distribution in the rolled strips can be seen in the Cumulative Distribution Function (CDF) plots in Figure \ref{fig:grain-size-weibull-area-frac} . Grain size was determined as the equivalent circle diameter from the EBSD data in Figure \ref{fig:rolled-ebsd} but excluding grains which intersect the micrograph boundary. \textcolor{black}{$\delta$-ferrite} grains were also excluded by first determining the percentage of $\delta$-ferrite grains out of the total ferrite grains in Table \ref{tab:tensile EBSD phase fraction}, \textit{i.e.} 16\% and 11\% of ferrite grains in the 5 kg (2 h) and 400 g (2 h) samples respectively. The ferrite grains were then sorted in decreasing size and the largest 16\% and 11\% of the ferrite grains were assumed to be $\delta$-ferrite and excluded since $\delta$-ferrite grains were consistently larger than $\alpha$-ferrite grains. From Figure \ref{fig:grain-size-weibull-area-frac}a, the austenite grain size CDFs were very similar. However, there was some spread in the $\alpha$-ferrite grains size CDFs between the three rolled strips. It is noteworthy that the average $\alpha$-ferrite grain size decreases slightly with increasing $\delta$-ferrite fraction. This effect may be attributed to a lower driving force for $\alpha$-ferrite grain growth during IA due to pre-existing $\delta$-ferrite grains. Nevertheless, Figure \ref{fig:grain-size-weibull-area-frac} shows that the grain size CDFs, especially of the austenite phase, are largely independent of casting condition and $\delta$-ferrite fraction.

\begin{table}[t]
	\centering
	\caption{EBSD phase fractions in \% and area weighted average grain size in $\mu$m of the rolled strips as determined by the Bruker ESPRIT software. N.I. $-$ Non indexed fraction.}
	\begin{tabular}{lccc|cc}
		\toprule
		& $\gamma$ & $\alpha$ & N.I. & d$_\gamma$ & d$_\alpha$ \\
		\midrule
		5 kg (2 h) & 54.3  & 44.1  & 1.6 & 3.6 & 4.0  \\
		400 g (2 h)  & 57.4  & 41.3   & 1.3  & 3.7 & 3.9  \\
		400 g (24 h)  &  62.4 & 36.9  & 0.7 & 3.4 & 2.4   \\
		\bottomrule
	\end{tabular}
	\label{tab:tensile EBSD phase fraction}%
\end{table}

From Figure \ref{fig:nov5a2-a3-wmg-24hr-2hr-tensiles}, the three rolled strips had nearly identical strain hardening behaviour regardless of casting condition or $\delta$-ferrite fraction. Since the austenite phase in all three strips had a similar grain size distribution (Figure \ref{fig:grain-size-weibull-area-frac}a), this then strongly implies that the austenite phase across all three samples should have the composition, \textit{i.e.} same Stacking Fault Energy (SFE) and stability (Md\textsubscript{30}). However, the austenite phase fraction was not constant across all three samples (Table \ref{tab:tensile EBSD phase fraction}), suggesting that the composition of the austenite could not have been identical under normal circumstances. Of the alloying additions in medium Mn steel, C is able to significantly alter the SFE and Md\textsubscript{30} of austenite, even in small concentrations \cite{Sun2018,Saeed-Akbari2009,Kwok2021}. Therefore, in order for the austenite phase in the partially homogenised samples to maintain the same C content as the fully homogenised sample while having a lower austenite fraction, it was likely that the excess C in the partially homogenised samples precipitated in the form of carbides along the $\delta$-ferrite interphase boundaries (Figure \ref{fig:rolled-ebsd}d). This phenomena was not observed in the 400 g (24 h) sample where there was no $\delta$-ferrite. The build up of C at $\delta$-ferrite interphase grain boundaries was also observed by other researchers \cite{Kim2019a,Sun2019}. 

%While the precipitation of carbides at $\delta$-ferrite interphase boundaries did not adversely affect the tensile properties, C segregation at interfaces is known to negatively affect impact properties in medium Mn steels \cite{Sun2019,Han2017}.

%The microstructures of the partially homogenised samples (Figure \ref{fig:rolled-ebsd}a-b) were very similar to the fully homogenised microstructure (Figure \ref{fig:rolled-ebsd}c) aside from the presence of several long ferrite grains. These long ferrite grains were most likely $\delta$-ferrite stringers which became elongated during rolling. In order to estimate the amount of $\delta$-ferrite in the rolled microstructure, the ferrite fraction of the 400 g (24 h) sample can be subtracted from the 5 kg (2 h) and the 400 g (2 h) samples. This results in an approximate $\delta$-ferrite fraction of 7.2\% and 4.4\% in the 5 kg (2 h) and 400 g (2 h) rolled strips respectively which is in close agreement with the $\delta$-ferrite fraction after the 2 h homogenisation heat treatment (Figure \ref{fig:delta-area-frac-graph}a). This suggests that any untransformed $\delta$-ferrite after homogenisation will be retained in the final microstructure.

\begin{figure}[t]
	\centering
	\includegraphics[width=\linewidth]{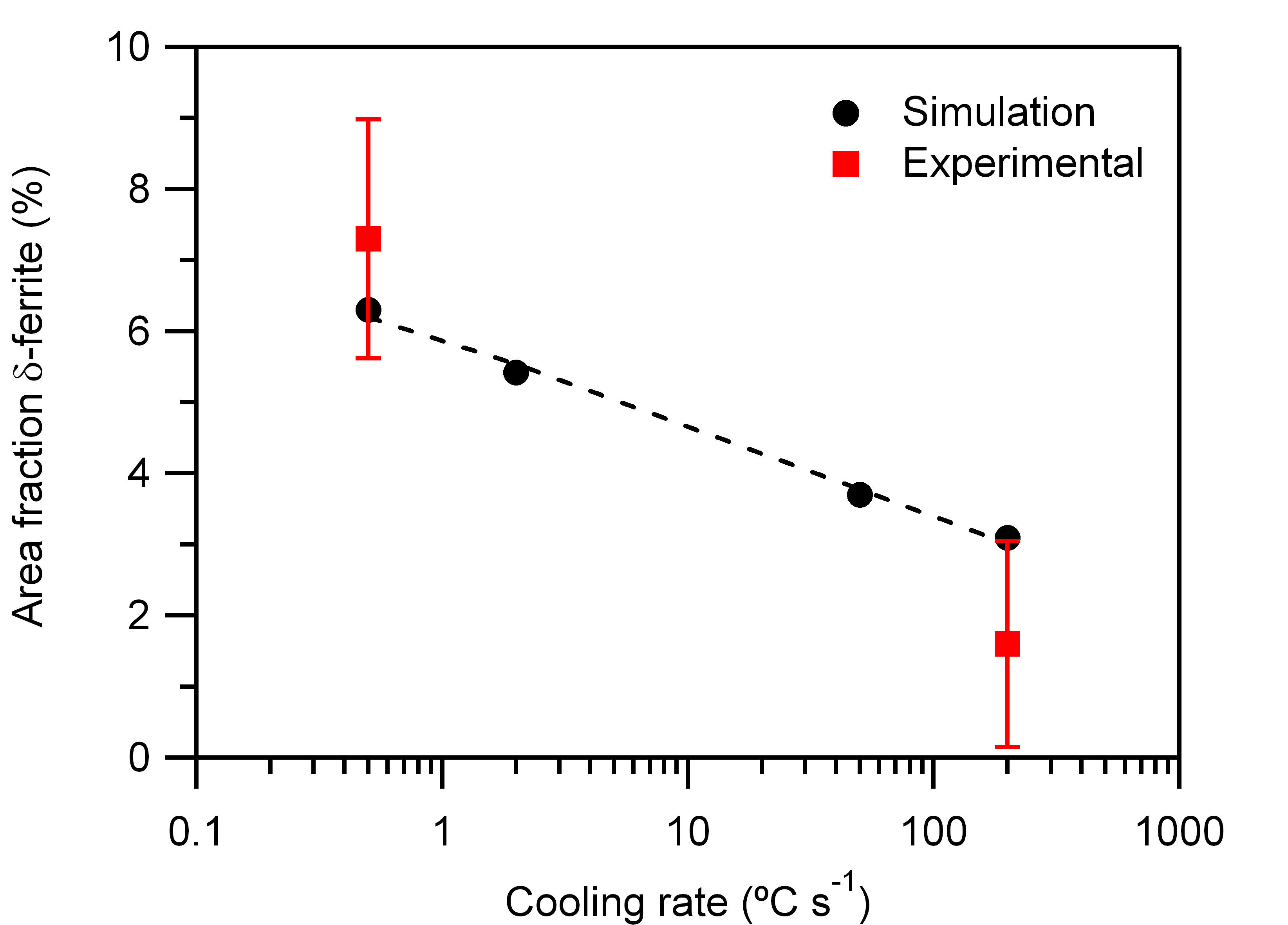}
	\caption{Variation of $\delta$-ferrite after 2 h homogenisation at 1250 \degree C with cooling rate. }
	\label{fig:area-frac-delta-after-hom-micress}
\end{figure}

From the post mortem 5 kg (2 h) tensile sample in Figure \ref{fig:rolled-ebsd}d, the fracture edge was observed to have sheared across the entire microstructure. Brittle transverse cleavage of $\delta$-ferrite grains was not observed near the fracture edge nor in the gauge. Near the fracture edge, voids grew around the $\delta$-ferrite interphase carbides but the voids remained very small and were not deemed to be a cause of fracture. Since the tensile samples which contained $\delta$-ferrite had slightly better elongation than the 400 g (24 h) sample, it was therefore unlikely that $\delta$-ferrite, at low fractions, would have a detrimental effect on elongation.

\subsection{Industrial relevance}

It has often been assumed but not studied in detail that medium Mn steels should be less prone to chemical segregation as compared to high Mn TWIP steels \cite{Wietbrock2010a,Reitz2011} on the basis of a significantly lower Mn content. While this was shown to be true in this study, the nature of segregation is also dependent on the process in which it was cast or produced. Since segregation is known to have a noticeable impact in medium and high Mn steels, the industrial production scale cast structure would have to be replicated at a laboratory scale for a true assessment of its real-world applicability. 

Two aspects of commercial production make lab scale feasibility trials difficult. Firstly, the level of rolling reduction. Commercial products are typically rolled from approximately 230 mm thick slab to 1 mm thick strip product, a reduction by more than 200 times which is difficult to attain in a lab. Secondly, the degree of chemical segregation in industrial sized slabs and the consequent effects on subsequent mechanical properties cannot be replicated through fast cooling laboratory small scale methods.

From Table \ref{tab:tensile EBSD phase fraction} and Figure \ref{fig:grain-size-weibull-area-frac}, it is evident that the scale of production has little impact on the final grain size and distribution, and as such, further rolling reduction during commercial production is not likely to refine this much more. However, this work has shown that there is a strong dependency on the casting parameters on the stability of $\delta$-ferrite and how it manifests in the final product. Figure \ref{fig:area-frac-delta-after-hom-micress} shows the influence of cast cooling rate on the area fraction of $\delta$-ferrite after 2 h homogenisation at 1250 \degree C. This difference has shown to have a noticeable impact on tensile properties with the slow cooled ingot showing a significant amount of $\delta$-ferrite that forms stringers after rolling. While a 24 h homogenisation of a laboratory scale 400 g ingot would be able to demonstrate the full capability of the steel, it does not truly represent the scalability of this product. Nevertheless, the 5 kg ingot gives confidence in the scalability of the production of this steel that still combines $>$1100 MPa tensile strength with $>$40\% elongation while only requiring a standard homogenisation cycle of 2 h at 1250 \degree C.

\section{Conclusion}

A medium Mn steel has been produced via two different processing routes. The emphasis of this work was to produce a steel with superior strength to elongation ratio while considering the full-scale production route to ensure true industrial relevance. Due to the nature of the chemical segregation of these samples, it is important that small scale production would be able to replicate some of the conditions in commercial production. This translation was aided by modelling both the solidification and homogenisation process in Micress. From this study, the following conclusions can be made:
\begin{enumerate}
	\item The as-cast microstructure depends heavily on the solidification rate of the ingot with the 400 g fast cooling sample showing significantly finer distribution of $\delta$-ferrite.
	\item The rate of homogenisation is also dependent on the initial cast microstructure. However, a 5 kg cast that represents the cooling rates seen in commericial production \textcolor{black}{showed} significant reduction in $\delta$-ferrite fraction even after a standard 2 h heat treatment. 
	\item Modelling has been carried out and good agreement with experiment can be seen. The modelling has revealed the trends in $\delta$-ferrite distribution and segregation for a range of casting rates \textcolor{black}{between 0.5 $-$ 200 \degree C \textsuperscript{-1}}.  
	\item Tensile testing has shown the excellent strength to elongation ratio of the steel. While the 5 kg ingot showed a slightly lower strength, the properties were still superior compared to current commercial steels.
	\item The decrease in strength in the 5 kg cast compared to the 400 g ingot can be attributed to the remnant $\delta$-ferrite. The coarser SDAS seen in the 5 kg ingot resulted in longer diffusion distances and therefore higher amount of $\delta$-ferrite can be seen in the final microstructure.
	\item The $\delta$-ferrite phase manifests itself as stringers in the final product. A sample of the 400 g ingot homogenised for 24 h then rolled and intercritical annealed showed no stringers and a random distribution of fine $\alpha$-ferrite, providing a more efficient source of strength to the steel.
\end{enumerate}

\section{Acknowledgements}
TWJK would like to acknowledge the provision of a studentship by A*STAR, Singapore. CS and CD would like to acknowledge funding support from UK EPSRC (Strategic equipment resource only grant, EP/V007548/1), Royal Academy of Engineering Support for Rapid Alloy Processing (REA1920/3/17) and High Value Manufacturing Catapult at WMG. XX and DD would like to acknowledge funding support from the UK EPSRC (Designing alloys for resource efficiency, EP/L025213/1).

\bibliographystyle{model1-num-names}
\bibliography{library.bib}

\end{document}